\documentclass[aip,apl,graphicx,preprint]{revtex4-1}
\usepackage{graphicx}
\usepackage{amsmath}
\usepackage{braket}

\begin{document}

%\preprint{}

%\title{Purcell Enhanced Quantum Dot Emission in Photonic Crystal Cavities Designed for Reduced Spectral Broadening} %Title of paper

\title{Multimode Nanobeam Photonic Crystal Cavities for Purcell Enhanced Quantum Dot Emission} %Title of paper



%\author{}
%\thanks{}
%\affiliation{}

\author{Junyeob Song}
\author{Ashish Chanana}
\affiliation{National Institute of Standards and Technology, Gaithersburg, MD, USA 20899}
\affiliation{Theiss Research, La Jolla, CA USA}
\author{Emerson G. Melo} 
\affiliation{University of S\~{a}o Paulo, Lorena, SP, Brazil}
\affiliation{National Institute of Standards and Technology, Gaithersburg, MD, USA 20899}
\author{William Eshbaugh}
\affiliation{Department of Physics and Astronomy, West Virginia University, 135 Willey St, Morgantown, West Virginia 26506, USA}
\affiliation{National Institute of Standards and Technology, Gaithersburg, MD, USA 20899}
\author{Craig Copeland}
\affiliation{National Institute of Standards and Technology, Gaithersburg, MD, USA 20899}
\author{Luca Sapienza}
\affiliation{Department of Engineering, University of Cambridge, Cambridge, CB3 0FA United Kingdom}
\author{Edward Flagg}
\affiliation{Department of Physics and Astronomy, West Virginia University, 135 Willey St, Morgantown, West Virginia 26506, USA}
\affiliation{National Institute of Standards and Technology, Gaithersburg, MD, USA 20899}
\author{Jin-Dong Song}
\affiliation{Center for Opto-Electronic Convergence Systems, Korea Institute of Science and Technology, Seoul 136-791, South Korea}
\author{Kartik Srinivasan}
\affiliation{Joint Quantum Institute, NIST/University of Maryland, College Park, MD, USA}
\affiliation{National Institute of Standards and Technology, Gaithersburg, MD, USA 20899}
\author{Marcelo Davanco}
\affiliation{National Institute of Standards and Technology, Gaithersburg, MD, USA 20899}
\email{junyeob.song@nist.gov}


\date{\today}

\begin{abstract}
    Epitaxial III-V semiconductor quantum dots in nanopthonic structures are promising candidates for implementing on-demand indistinguishable single-photon emission in integrated quantum photonic circuits. Quantum dot proximity to the etched sidewalls of hosting nanophotonic structures, however, has been shown to induce linewidth broadening of excitonic transitions, which limits emitted single-photon indistinguishability. Here, we design and demonstrate GaAs photonic crystal nanobeam cavities that maximize quantum dot distances to etched sidewalls beyond an empirically determined minimum that curtails spectral broadening. Although such geometric constraint necessarily leads to multimode propagation in nanobeams, which significantly complicates high quality factor cavity design, we achieve resonances with quality factors $Q\approx10^3$, which offer the potential for achieving Purcell radiative rate enhancements $F_p\approx100$. 
\end{abstract}

\pacs{}% insert suggested PACS numbers in braces on next line

\maketitle %\maketitle must follow title, authors, abstract and \pacs

%\section{}
%\label{}

On-chip sources of indistinguishable on-demand single-photons are highly desirable for integrated photonic quantum technologies. Single self-assembled InAs epitaxial quantum dots (QDs) have demonstrated near-ideal single-photon emission at rates exceeding $10^9$~s$^{-1}$ when integrated into GaAs nanophotonic geometries designed to efficiently funnel emitted photons into desirable optical spatial modes~\cite{Davanco2011, Claudon2010, somaschi_near-optimal_2016, uppu_scalable_2020, senellart_high-performance_2017, rickert_high_2025}. 

Integration into nanophotonic cavities allows reduction of QD radiative transition lifetimes ($T_1$) via the Purcell effect, which translates to higher achievable single-photon trigger rates~\cite{rickert_high_2025}. In addition, an increased radiative rate brings the coherence time, $T_2$, closer to the Fourier limit $T_2=2T_1$, which is necessary for high indistinguishability~\cite{Liu2018,rickert_high_2025}. The difference between $T_2$ and the Fourier limit is primarily determined by phonon dephasing and fluctuating local electric and magnetic fields~\cite{Kuhlmann2013a}. Physically, QD spectral fluctuations at time scales comparable to $T_1$ can be circumvented with faster radiative decay~\cite{Liu2018,rickert_high_2025}. Liu {\it et al.}, have shown, however, that an etched GaAs surface in close proximity ($< 300$ nm) to a single QD leads to significant spectral broadening of the emission from the same QD transitions, measured before and after fabrication~\cite{Liu2018a}. The observed broadening is most likely spectral diffusion arising from charge fluctuations (on time scales $\gg T_1$) of surface or defect states created by etching. While the incorporation of QDs into p-i-n heterostructures~\cite{Thyrrestrup2018} and potentially the use of different surface passivation techniques~\cite{Liu2018a,manna_surface_2020} can mitigate such effects, another approach is to develop photonic geometries that maintain a minimum separation of 300 nm between the QD and etched features. In the design of nanophotonic cavity modes with high Purcell factors ($F_p \sim Q/V$ where $Q$ and $V$ are the quality factor and modal volume, respectively), achieving small $V$ while maintaining sufficiently large distances between etched sidewalls and the QD is challenging~\cite{melo_multiobjective_2023}. In particular, in the air-clad, free-standing GaAs thin film platform with typical thickness of about 200~nm utilized for GaAs photonic nanostructures~\cite{Davanco2011}, a waveguide of width $\geq600$~nm supports multiple waveguide modes at wavelengths near 940~nm, to which embedded InAs QDs can efficiently couple. The presence of more than one waveguide mode complicates extraction of light into a single mode, which would be necessary for downstream light guiding and manipulation.

In this study, we show that photonic crystal (PhC) cavities consisting of etched hole arrays in wide, GaAs multimode nanobeam (MMNB) waveguides can be designed to achieve significant Purcell radiative rate enhancement for InAs QDs embedded at distances greater than 300~nm from etched surfaces.

\begin{figure}[hbt!]
    \centering 
    \includegraphics[width=0.9\textwidth]{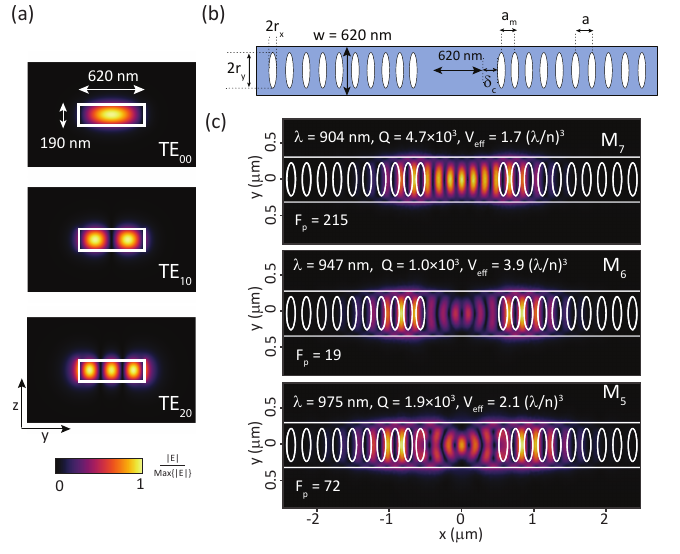}
    \caption{(a) Supported Transverse-Electric (TE) modes of an air-clad multimode nanobeam (MMNB) waveguide with a width of 620~nm and thickness of 190~nm at a wavelength of 940~nm. (b) Schematic of the photonic crystal (PhC) Fabry-Perot type cavity implemented on a MMNB, composed of a lattice of elliptical holes with major and minor axis radii $r_x$ and $r_y$ respectively. The hole spacing varies smoothly from $a_m$ closest to the cavity center to a nominal value $a$, as described in the text. Importantly, a QD located at the cavity center is located 310~nm away from the NB edges, and 620 nm + $2\cdot\delta_c$ from the first holes. (c) Electric field distributions for modes $M_5$, $M_6$ and $M_7$ supported by a representative MMNB cavity (geometrical parameters specified in the main text), obtained by finite element method calculations. Resonance wavelengths ($\lambda$), quality factors ($Q$), mode volumes ($V_{\mathrm{eff}}$) and Purcell factors ($F_p$) at the field maximum within the cavity center are displayed. $n\approx3.49$ is the GaAs refractive index.}
    \label{Fig1}
\end{figure}

The MMNB waveguide onto which the PhC cavity is implemented has a thickness of 190~nm and a width of 620~nm, thereby supporting three transverse-electric (TE) guided modes within the range of quantum dot emission centered around 940~nm. Field profiles for such modes, labeled TE$_{00}$, TE$_{10}$ and TE$_{20}$, are shown in Fig.~\ref{Fig1}(a). The cavity consists of two etched, one-dimensional PhC mirrors with elliptical holes and an unetched central confinement region, forming a Fabry-Perot resonator as shown in Fig.~\ref{Fig1}(b). The next-neighbor hole spacing $a_n$ for the first six holes of the PhC mirror follow a smooth function from the center, $a_{n+1}=a+2^{-0.5(n/\delta_a)^2}(a_m-a)$, with $n\geq0$. In this expression $n$ are hole indices, $a_m$ the minimum spacing nearest the cavity center, $a$ the maximum (nominal) spacing, and $\delta_a$ a dimensionless transition steepness. For $n>6$, $a_n=a$. The elliptical hole radii, $r_x$ and $r_y$, are constant along the nanobeam for the geometry in Fig.~\ref{Fig1}(b).

The multimode character of the nanobeam and the wide unetched central region complicate the application of deterministic cavity design~\cite{Loncar_nanobeam_2011}. To overcome this challenge, we performed a scan of the geometrical parameters indicated in Fig.~\ref{Fig1}(b) to maximize the Purcell factor $F_P$ experienced by an electric dipole at the cavity center. Importantly, all parameters are constrained to ensure the central unetched region is at least 620~nm wide in all directions. 
Values of $F_p$  were determined through 3D finite-difference time-domain (FDTD) simulations. In these simulations, a y-oriented radiating electric dipole, representing a single QD, was positioned at the center of the cavity, and the ratio $P_{rad}/P_{hom}=F_p$ was calculated, which represents the ratio between the total steady-state power emitted in the cavity ($P_{rad}$) and in a homogeneous medium ($P_{hom}$). 

Fig.~\ref{Fig1}(c) shows a representative cavity geometry, resulting from the parameter search, that supports three TE resonances within a wavelength range of 900~nm to 1000~nm, with quality $Q$ factors in the $10^3$ range, as indicated. Here, $a=203$~nm, $a_m=171$~nm, $\delta_c=191$~nm and $\delta_a=1.45$. The short-, intermediate- and long-wavelength resonances, respectively, feature seven, six and five electric field anti-nodes along $x$ in the cavity region, and are labeled $M_7$, $M_6$, and $M_5$. Modes $M_7$ and $M_5$ are symmetric across $x=0$, and $M_6$ is anti-symmetric. The transverse field profile of $M_7$ suggests a large TE$_{00}$ waveguide mode component to its constitution. In contrast, the field profiles for modes $M_5$ and $M_6$ are reminiscent of multimode field beating along $x$, suggesting a more significant TE$_{20}$ content. 

We calculated Purcell factors for y-oriented dipole emitters located at the central antinodes of the three modes via $F_p=(3/4\pi^2)\cdot Q/V_{\mathrm{eff}}$, where $V_{\mathrm{eff}}=V/(\lambda/n)^3$ is the effective mode volume, $V=\int{\epsilon(\mathbf{r})\left|\mathbf{E}(\mathbf{r})\right|^2dV}/\epsilon(\mathbf{r}_0)\left|\mathbf{E}(\mathbf{r}_0)\right|^2$, $n\approx3.49$ is the GaAs refractive index, and $\mathbf{r}_0$ is the dipole location in the cavity. For $M_5$ and $M_7$, $\mathbf{r}_0$ was at the cavity center, whereas for $M_6$ it was located at the first off-center field anti-node. Calculated values for $V_{\mathrm{eff}}$ and $F_p$ are displayed in Fig.~\ref{Fig1}(c), where is is apparent that relatively small mode volumes can be achieved, contributing to reasonably high Purcell factors.

Several cavities as in Fig.~\ref{Fig1}(b), with small parameter variations, were fabricated on an epitaxially grown wafer consisting of a 190~nm thick layer of GaAs, which contained a high ($>1~\mu \mathrm{m}^{-2}$) density of InAs QDs at the center. The GaAs layer was grown on top of a sacrificial AlGaAs layer with a thickness of 1~$\mu$m. The fabrication process involved electron-beam lithography and inductively-coupled plasma reactive-ion etching to transfer the nanobeam patterns onto the GaAs layer containing the QDs. Subsequently, the AlGaAs sacrificial layer was removed using hydrofluoric acid.
\begin{figure}[hbt!]
    \centering
    \includegraphics[width=0.6\textwidth]{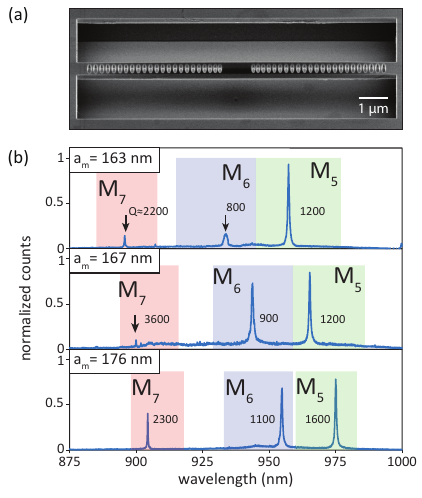}
    \caption{Characterization of fabricated GaAs MMNB waveguides. (a) Scanning electron micrograph of a fabricated MMNB PhC. (b) Measured photoluminescence spectra, originating from an ensemble of quantum dots under strong non-resonant optical excitation, for MMNB cavities with nominal $a$, $r_x$, $r_y$, $\delta_a$ and $\delta_c$ as in Fig.~\ref{Fig1}(c) and varying nominal minimum lattice constant parameter $a_m$. Red, blue and green shaded areas correspond to spectral ranges of modes M$_7$, M$_6$ and M$_5$, calculated with the finite element method, assuming enlargement of nominal PhC hole diameters by as much as 6~nm, as well as of refractive index variations (based on \textcite{skauli_improved_2003}) over the displayed wavelength range. Sharp resonances correspond to cavity resonances with approximate quality factors Q as indicated.
    }
    \label{Fig2}
\end{figure}
Figure~\ref{Fig2}(a) displays a scanning electron micrograph of a representative fabricated device. We used a micro-photoluminescence ($\mu$PL) imaging and spectroscopy setup to characterize QD emission in the fabricated devices, at a temperature of 1.8~K. The devices were illuminated with a continuous-wave (CW), 780 nm laser spot with a diameter of approximately 1 $\mu$m, and QD-emitted light collected from a confocal spot was coupled into a single-mode fiber and dispersed in a grating spectrometer. The excitation power level was increased until sharp resonances were clearly visible in the PL spectrum. Figure~\ref{Fig2}(b) shows the PL spectra of the QD ensemble in three fabricated devices with varying PhC mirror lattice parameter $a_m$, as indicated in the caption. The spectra are representative for the tested MMNB cavities, and each one features three prominent peaks that correspond to different optical resonances, with approximate quality factors as indicated in the figure. Here, the 780 nm excitation, above the GaAs bandgap, generates carriers in the GaAs host that eventually populate the QDs, and excitonic QD emission feeds the cavity modes. It is worth noting that the QD ensemble emission spectrum measured at regions away from the cavities extends between approximately 900 nm and 1050 nm with no sharp resonances such as seen in Fig.~\ref{Fig2}(b).   

To make modal assignments to the experimentally observed PL peaks, we proceeded as follows. Using a finite element model, we determined the possible wavelength ranges for modes $M_7$,  $M_6$ and  $M_5$ (see Fig.~\ref{Fig1}(c)), considering reasonable, potential geometrical and refractive index variations. Specifically, cavity geometries with dimensions taken from scanning electron micrographs (SEM) of fabricated cavities were used, and hole dimensions were allowed to vary by up to 6~nm. In addition, temperature- and wavelength-dependent GaAs refractive index models~\cite{papatryfonos_refractive_2021, skauli_improved_2003} were used to obtain maximum and minimum refractive index ranges for the three modes. In Fig.~\ref{Fig2}, the red, blue and green shaded areas correspond to spectral ranges where $M_7$, $M_6$ and $M_5$, respectively were found in simulation, considering those variations. We note that the quality factors for the different modes did not change considerably with the applied geometrical or index variations. 

\begin{figure}[hbt!]
    \centering
    \includegraphics[width=0.8\textwidth]{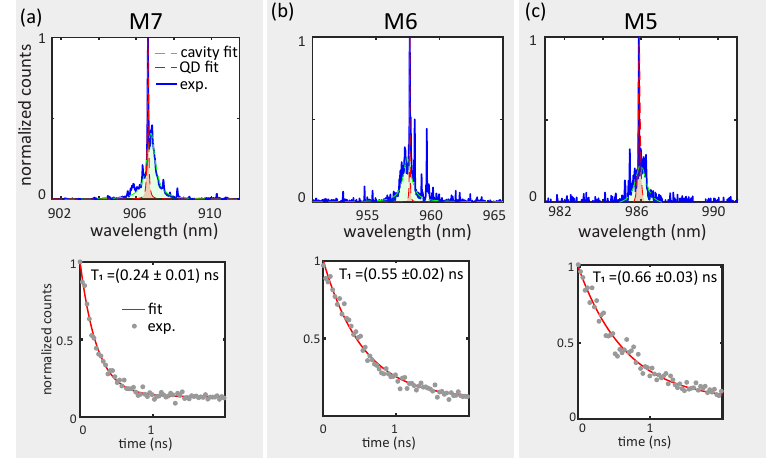}
    \caption{(a)-(c): Emission spectra and exciton radiative decay traces for representative quantum dots in M7, M6 and M5 cavity resonances, respectively. Red and green dashed lines and shades are least-square fits to the QD exciton and cavity, respectively. Corresponding exciton decay traces are also shown, fitted with single exponential decay curves. Uncertainties are 95~\% confidence intervals (two standard deviations) from the fits.}
    \label{Fig3}
\end{figure}
Using low power, non-resonant excitation at 780~nm, we located and identified the emission lines from single QD excitons coupled to the $M_7$, $M_6$ and $M_5$ resonances of various devices. Figures~\ref{Fig3}(a)-(c) show representative spectra observed from $M_7$, $M_6$ and $M_5$ resonances of different fabricated cavities. The higher intensity peaks are exciton transitions from single QDs that are spectrally aligned with a broader resonance corresponding to the cavity. Pumping with a 76~MHz train of approximately 140~fs, 806~nm pulses, we performed radiative decay measurements of the cavity-coupled QD excitons, using a silicon single-photon avalanche detector (SPAD) and a time-correlated single-photon counting module, with a timing resolution of approximately 80~ps. Representative decay traces corresponding to the respective spectra are also shown in Figs.~\ref{Fig3}(a)-(c). The decay traces are fitted with a single exponential decay function with decay times as indicated in the figures. We measured eleven single quantum dot exciton lines coupled to $M_5$, $M_6$ or $M_7$ modes of eleven different fabricated cavities. Overall, considering a natural lifetime of 1~ns, which is typical for a QD transition in bulk GaAs, Purcell factors of less than 5 were observed (see Table S1 in the SI)---considerably lower than predicted by our models.

We postulate a combination of two mechanisms to explain this observation. First, lack of spatial location, orientation, and spectral control of single QDs within the cavities individually can lead to significant departure from predicted modal coupling and Purcell factors~\cite{copeland_traceable_2024}. Second, under non-resonant excitation, the slow carrier relaxation time that precedes the QD exciton formation and emission leads to significantly longer observed radiative decay traces. Indeed, ~\textcite{Liu2018} have reported decay lifetimes for above-band QD excitation that were more than ten times longer than those measured with a resonant excitation measurement that avoided the carrier relaxation bottleneck. As we show in the SI, a three-level system rate equation model involving non-radiative relaxation from a high energy state into a lower, Purcell-enhanced radiative state predicts radiative decays with lifetimes that are limited by either the carrier relaxation time or radiative decay time. Considering carrier relaxation times from 100~ps to 1000~ps, consistent with ~\textcite{reithmaier_carrier_2014}, Fig. S1 in the SI predicts significantly lower observed Purcell factors than expected from MMNB simulations.

To further test such explanation, we employed a Monte Carlo approach to randomly sample QD positions and orientations within each of the three cavity modes, determining distributions of expected Purcell enhancements for a natural QD lifetime of 1.0~ns and considered only the lifetime of the neutral exciton state. We then accounted for the effects of spectral misalignment and experimental versus theoretical cavity Q, producing Purcell factor distributions for each of our experimental devices. The three-level system model described above then provides relationships for estimating the observable Purcell factor resulting from non-resonant excitation as a function of the Purcell factor resulting from only the lifetime of the neutral exciton state, given the carrier relaxation time (Fig. S1 in the SI). The final results are Purcell factor distributions for each device for carrier relaxation times ranging from 100 ps to 500 ps (Fig. S2 in the SI). These indicate that observation of Purcell factors greater than 10 is highly unlikely, consistent with our experimental results. However, questions remain about the frequency of the observation of such low Purcell factors, which appear consistent with carrier relaxation times in the range of 100~ps to 500~ps, and will be the subject of further investigations.

In summary, our work shows that GaAs nanobeams with widths of more than 600~nm can be leveraged for the creation of photonic crystal cavities capable of supporting optical resonances with Q in the $10^3$ range at wavelengths of approximately 904~nm. A width of 600 nm ensures that QDs can be at least 300~nm away from etched sidewalls, a minimum to avoid excess QD spectral diffusion as observed in \textcite{Liu2018a}, without relying on surface passivation~\cite{manna_surface_2020} or static electric field application~\cite{Thyrrestrup2018}. Importantly, a width of 600~nm means that the nanobeam itself supports multiple guided TE modes, a fact that imposes challenges for achieving high quality resonances by known deterministic design methods~\cite{Loncar_nanobeam_2011}, since such methods target single-mode nanobeams. Simulations indicate that our multimode nanobeam PhC cavities can nonetheless support large ($F_p > 200$) Purcell radiative rate enhancements on embedded quantum dots. The use of MMNB cavities for waveguided resonant excitation via high-order nanobeam modes is anticipated and is the subject of another study~\cite{chanana_modular_2025}. Experimentally fabricated cavities have shown cavity resonances within the predicted wavelength ranges, with comparable quality factors. While experimentally estimated Purcell factors are considerably lower than predicted, we anticipate that deterministic QD positioning~\cite{Sapienza2015,copeland_traceable_2024} to maximize QD-cavity coupling, and resonant excitation~\cite{Liu2018} to avoid slow carrier relaxation effects, will allow proper verification of the effect. Our work indicates a viable strategy to circumvent an important hurdle in the creation of highly indistinguishable on-chip single-photon sources based on single epitaxial quantum dots.

% If in two-column mode, this environment will change to single-column format so that long equations can be displayed. 
%\begin{widetext}
%$$\mbox{put long equation here}$$
%\end{widetext}

% Figures should be put into the text as floats. 
% Use the graphics or graphicx packages (distributed with LaTeX2e).
% Put the label that you will use with \ref{} command in the braces of the \label{} command.
%
% \begin{figure}
% \end{figure}

% \begin{table}
% \end{tabular}
% \end{table}

\begin{acknowledgments}
This work was performed under the financial assistance award 70NANB21H059 from the U.S. Department of Commerce, National Institute of Standards and Technology. We thank Thiago P. M. Alegre from the Univeristy of Campinas, Brazil, for very helpful discussions. W. Eshbaugh acknowledges financial assistance under award 70NANB23H257 from U.S. Department of Commerce, National Institute of Standards and Technology. E.B. Flagg acknowledges support from the Office of Naval Research (ONR) under award number N00014-23-1-2611. JDS acknowledges the partial support from KIST Institution Program (2E32942).
% Put your acknowledgments here.
\end{acknowledgments}

\bibliographystyle{apsrev4-1}
\bibliography{references}

\clearpage
\onecolumngrid
\begin{center}
    \large\bfseries Supplementary Information\\[1ex]
    \normalsize Multimode Nanobeam Photonic Crystal Cavities for Purcell Enhanced Quantum Dot Emission
\end{center}
\vspace{1em}
\date{\today}

\section{ Carrier relaxation time effect on observed excited state lifetimes }
\label{section:carrier_relaxation}
\begin{figure}[h]
    % \centering
    \includegraphics[width=0.8\linewidth]{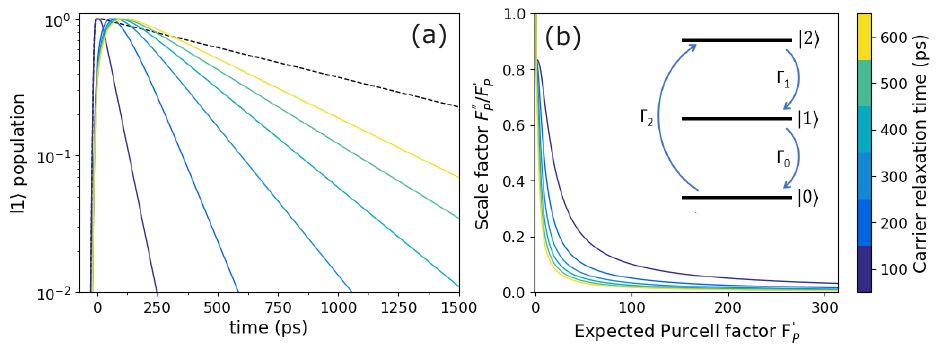}
    \caption{
    3LS modeling of above-band excitation in a quantum dot. (a) $p_1$ population dynamics for a range of carrier relaxation times using the same input exciton decay rate $\Gamma_0=100$ ps$^{-1}$. The dashed line represents a 1~ns lifetime seen in typical bulk QDs using a negligible carrier relaxation time. (b) Ratio of observed ($F_P^{''}$) and expected ($F_P^{'}$) Purcell factors . Using a reference lifetime of 1~ns, $F_P^{''}$ is calculated from lifetimes extracted from exponential decay fits of the populations shown in (a), while $F_P^{'}$ is the Purcell enhancement factor for $\Gamma_0$, e.g. $F_P^{'}=\Gamma_0/\Gamma_\text{hom.}$, with $\Gamma_\text{hom.}$ is the dipole emission rate in bulk GaAs. (inset) Energy level diagram depicting the ground state $\ket{0}$, exciton $\ket{1}$, and higher-energy state $\ket{2}$. $\Gamma_2$ is the above-band excitation rate, $\Gamma_1$ is the non-radiative carrier relaxation rate, and $\Gamma_0$ is the exciton decay rate.
    }    
    \label{3LS}
\end{figure}

A simple three-level-system (3LS) model is employed to explain the lifetime-limiting effect of carrier relaxation time. For incoherent excitation, the system can be described by a set of coupled rate equations,
\begin{equation} \label{rate_eqn}
    \frac{d}{dt} 
    \begin{bmatrix}
    p_0\\
    p_1\\
    p_2\\
    \end{bmatrix}
    = 
    \begin{bmatrix}
    -\Gamma_2e^{-t^2/2\tau} & \Gamma_0 & 0\\
    0 & -\Gamma_0 & \Gamma_1\\
    \Gamma_2e^{-t^2/2\tau} & 0 & -\Gamma_1
    \end{bmatrix}
    \begin{bmatrix}
    p_0\\
    p_1\\
    p_2\\
    \end{bmatrix}.
\end{equation}
Shown in the inset of Figure~\ref{3LS}(b), $p_0$, $p_1$, and $p_2$ are the populations of the ground, intermediate-energy, and higher-energy states, respectively. $\Gamma_0$, $\Gamma_1$, and $\Gamma_2$ are the transition rates to the subscript state. The exponential terms in Eq.~\ref{rate_eqn} describe the Gaussian excitation pulse with temporal width $\tau$. Population dynamics are obtained by solving Eqn.~(\ref{rate_eqn}) numerically. The results for $p_1(t)$ are shown in Figure~\ref{3LS} below.

\section{Monte Carlo simulations of observed Purcell factors}
\label{section:monte_carlo}
We combine the effects of carrier relaxation due to non-resonant excitation with a Monte Carlo model of QD position and orientation to determine theoretical distributions of observable Purcell factor for each experimental device. The model begins with randomly sampling uniform distributions for position (x,y) that cover the area of the beam center between the first set of elliptical holes and within a minimum distance of 40~nm from device edges. For each position, we then randomly sample a dipole orientation $\theta$ from a uniform distribution over the range $[0, \pi/2]$, and assign an additional orthogonal dipole $\theta+\pi/2$. We do this to account for the two orthogonally polarized fine-structure split neutral exciton transitions, which we assume to be equally likely upon non-resonant excitation. For each sample, the resulting Purcell factor for the neutral exciton state of an ideal nanobeam is then given by
\begin{equation}
F_p(x,y,\theta)=\frac{|E_x(x,y)\cos\theta|^2+|E_y(x,y)\sin\theta|^2}{E_\text{max}^2} F_p^\text{max},
\end{equation}
where the electric fields $E_x$, $E_y$ are given by the FDTD simulation results for the different modes in Fig.~1(c) of the main text, $E_\text{max}$ is the maximum of the electric field, and $F_p^\text{max}$ is the maximum Purcell factor. We consider that there is an equal probability to observe emission from either the $\theta$ dipole or the $\theta+\pi/2$ dipole, and we draw $10^6$ values with equal probability from distributions for $F_p (x,y,\theta)$ and $F_p (x,y,\theta+\pi⁄2)$.
We then apply two multiplicative factors that reduce the theoretical Purcell factor to account for any difference between the experimental and theoretical cavity Q, and to account for spectral misalignment between the QD wavelength, $\lambda$, and cavity resonance wavelength, $\lambda_r$. Figure 1(c) in the main text shows these theoretical values for each mode. The reduced Purcell factor is
\begin{equation}
    F_p'(x,y,\theta)=\frac{Q_\text{exp}}{Q} \left[1+4Q_\text{exp}^2 \left(\frac{\lambda_r}{\lambda}-1\right)^2\right]^{-1} F_p (x,y,\theta).
    \label{eq:Fp_prime}
\end{equation}

\begin{figure}[h]
    % \centering
    \includegraphics[width=0.9\linewidth]{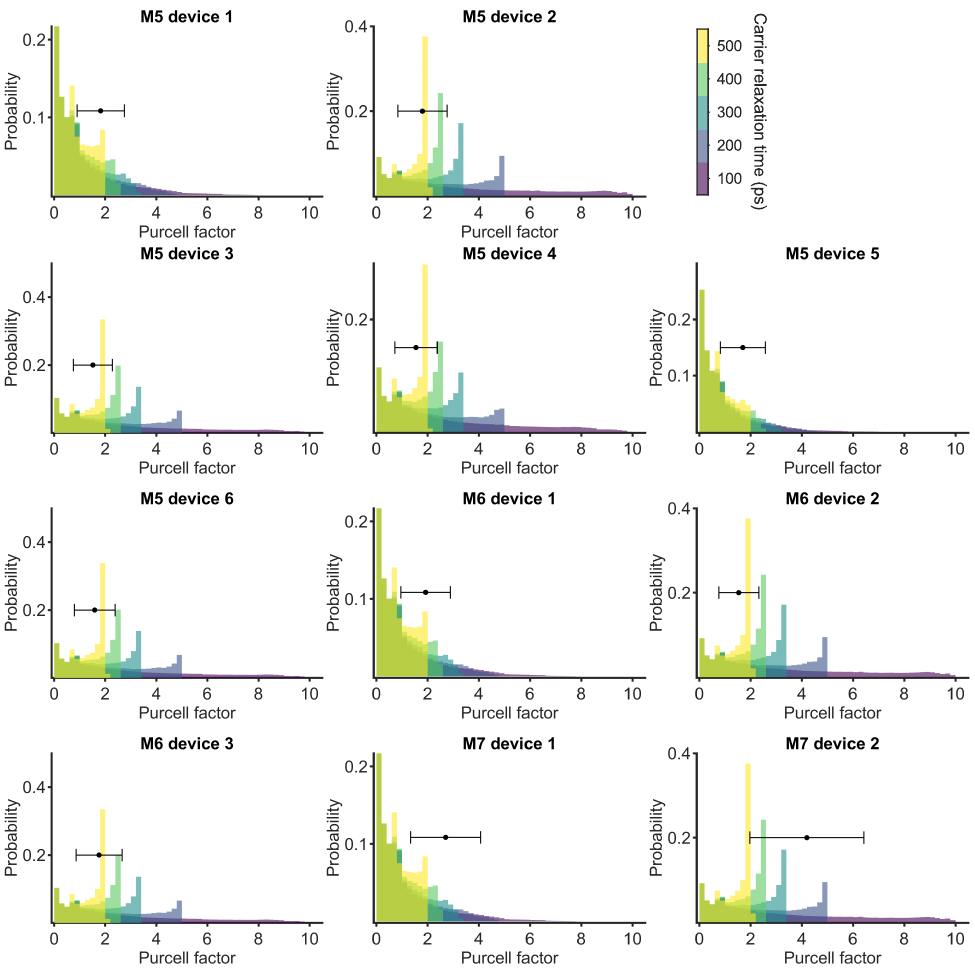}
    \caption{Theoretical estimates of observable Purcell factor. Probability histograms of observable Purcell factor for 11 experimental devices. Data markers indicate experimental values. Uncertainties are 95~\% confidence intervals.
    }
    \label{Fig:histograms}
\end{figure}

The three-level system model described in Section~\ref{section:carrier_relaxation} above then provides relationships for estimating the observable Purcell factor resulting from non-resonant excitation, $F_p''$, as a function of the Purcell factor resulting from only the lifetime of the neutral exciton state, $F_p'$, given the carrier relaxation time (Fig.~\ref{3LS}). Figure~\ref{Fig:histograms} shows the final distributions of observable Purcell factor for each of the 11 experimental devices.

% \begin{figure}[h]
%     % \centering
%     \includegraphics[width=0.95\linewidth]{Figure_S-2.pdf}
%     \caption{Reduction of observable Purcell factor due to carrier relaxation from non-resonant excitation. Plot showing the ratio of $F_p''$ and $F_p'$ as a function of $F_p'$ for different carrier relaxation times.}
%     \label{Fig:Fp_reduction}
% \end{figure}

\clearpage
\section{Experimental cavity-QD coupling parameters}
Table~\ref{table:cavity_QD_parms} shows experimentally observed coupling parameters for single quantum dot excitons in resonances of various fabricated cavities. We used such parameters (quality factor $Q_\text{exp}$; cavity wavelength, $\lambda$; QD-cavity spectral detuning, $\lambda-\lambda_r$), with Eq.~(\ref{eq:Fp_prime}) for the Monte-Carlo simulations described in the SI Section~\ref{section:monte_carlo}. Corresponding single QD exciton decay lifetimes obtained from fits to measured decay traces are also displayed. The uncertainites for the cavity center wavelengths ($\lambda_r$) and quality factors ($Q_\text{exp}$) were less than 1~\%. Other reported uncertainties were obtained from fits to experimental data, and correspond to 95~\% fit confidence intervals.
\begin{table}[h]
\caption{Experimental cavity-QD coupling parameters.}
\begin{tabular}{|c|c|c|c|c|}
\hline
Resonance and device number & $Q_\text{exp} (\times 10^3)$ & $\lambda$~(nm) & $\lambda_r-\lambda$~ (nm) & Lifetime (ns)    \\ \hline
M7, device 1        & 1.6                          & 907            & $0.18 \pm 0.04$         & $0.24 \pm 0.04$  \\ \hline
M7, device 2        & 0.9                          & 894            & $0.39 \pm 0.05$          & $0.37 \pm 0.02$  \\ \hline
M6, device 1        & 0.9                          & 958            & $0.09 \pm 0.05$          & $0.52 \pm0.03$   \\ \hline
M6, device 2        & 1.0                          & 966            & $0.27 \pm 0.05$          & $0.65 \pm 0.05$  \\ \hline
M6, device 3        & 1.0                          & 969            & $0.17 \pm 0.07$          & $0.57 \pm 0.06$ \\ \hline
M5, device 1        & 1.0                          & 975            & $0.03 \pm 0.07$          & $0.65 \pm 0.11$  \\ \hline
M5, device 2        & 1.1                          & 975            & $0.71 \pm 0.08$          & $0.59 \pm 0.07$ \\ \hline
M5, device 3        & 1.4                          & 986            & $0.12 \pm 0.04$          & $0.66 \pm 0.05$  \\ \hline
M5, device 4        & 1.2                          & 974            & $0.03 \pm 0.02$         & $0.63 \pm 0.03$  \\ \hline
M5, device 5        & 1.2                          & 988            & $0.6 \pm 0.02$           & $0.55 \pm 0.04$  \\ \hline
M5, device 6        & 1.5                          & 984            & $0.07 \pm 0.02$          & $0.56 \pm 0.1$   \\ \hline
\end{tabular}
\label{table:cavity_QD_parms}
\end{table}

\end{document}